\documentclass[reprint, superscriptaddress]{revtex4-1}

\usepackage[pdftex]{graphicx}
\usepackage{color}
\usepackage{amsmath}
\usepackage{upgreek}
\begin{document}

\title{Robust High-Dynamic-Range Vector Magnetometry via Nitrogen-Vacancy Centers in Diamond}

\author{Hannah Clevenson}
\email{These authors contributed equally to this work.}
\affiliation{Massachusetts Institute of Technology, Cambridge, MA 02139, USA}
\affiliation{MIT Lincoln Laboratory, Lexington, MA 02420, USA}

\author{Linh M. Pham}
\email{These authors contributed equally to this work.}
\affiliation{MIT Lincoln Laboratory, Lexington, MA 02420, USA}

\author{Carson Teale}
\email{These authors contributed equally to this work.}
\affiliation{Massachusetts Institute of Technology, Cambridge, MA 02139, USA}
\affiliation{MIT Lincoln Laboratory, Lexington, MA 02420, USA}

\author{Kerry Johnson}
\affiliation{MIT Lincoln Laboratory, Lexington, MA 02420, USA}

\author{Dirk Englund}
\affiliation{Massachusetts Institute of Technology, Cambridge, MA 02139, USA}

\author{Danielle Braje}
\email{braje@ll.mit.edu}
\affiliation{MIT Lincoln Laboratory, Lexington, MA 02420, USA}

\date{\today}

\begin{abstract}

We demonstrate a robust, scale-factor-free vector magnetometer, which uses a closed-loop frequency-locking scheme to simultaneously track Zeeman-split resonance pairs of nitrogen-vacancy (NV) centers in diamond.
Compared with open-loop methodologies, this technique is robust against fluctuations in temperature, resonance linewidth, and contrast; offers a three-order-of-magnitude increase in dynamic range; and allows for simultaneous interrogation of multiple transition frequencies. By directly detecting the resonance frequencies of NV centers aligned along each of the diamond's four tetrahedral crystallographic axes, we perform full vector reconstruction of an applied magnetic field.
\end{abstract}

%\pacs{}
\maketitle

\section{Introduction}

Quantum magnetometers have seen rapid advancement in the past two decades, with total-field sensitivities now rivaling those of cryogenic SQUIDs~\cite{Aleksandrov1995, BudkerPRA2000, KominisNature2003}. The nitrogen-vacancy (NV) color center in diamond is a particularly promising platform for a new class of quantum magnetic sensor that combines the long-term stability of an atom-like system with inherent vector capability in a compact, solid-state package. Similar to atomic magnetometers~\cite{budker2013optical}, the NV center ties external field measurements to fundamental constants through Zeeman splitting of quantized energy levels~\cite{Taylor2008,Maze2008,Balasubramanian2008}. The solid-state platform facilitates the incorporation of a high density of NV centers (up to $10^{18}\ \textrm{cm}^{-3}$~\cite{Acosta2009,Acosta2010a,Barry2016}) into a compact sensor volume. Furthermore, the tetrahedral structure of the diamond crystal lattice allows such devices to directly sense the magnetic field vector~\cite{SteinertRSI2010, MaertzAPL2010, PhamNJP2011}, an advantage over the inherently scalar measurements of traditional atomic magnetometers. The combination of these characteristics in a single magnetic sensor is transformational for a wide variety of magnetometry applications, ranging from all-magnetic navigation \cite{Cui201570, Shockley:2014uq, CancianiIEEE2017} to geological surveys \cite{foote1996relationship} to magnetic mapping of archaeological sites \cite{kvamme2006magnetometry} to locating unexploded ordinance \cite{espy2010ultra, sudac2009detection, mcfee1994total}. In this Article, we overcome a number of major technical obstacles to make significant progress towards realizing robust, fieldable NV magnetometers, suitable for operation in the practical environments required for many of these potential applications.

In this work, we employ a frequency-locking magnetic sensing scheme~\cite{SchoenfeldPRL2011,GrinoldsNatNano2014} to directly measure the NV resonance frequencies. Using custom-built electronics to address both Zeeman-split resonances simultaneously, we decouple magnetic field measurements from temperature, important for field applications where careful temperature control is not plausible. We show the immunity of this scale-factor-free measurement technique to variable parameters such as microwave and laser intensity noise and perform real-time magnetic sensing with high dynamic range. This capability is particularly necessary for non-stationary vector magnetic sensors, where rotation in an ambient field (e.g., Earth's field) may sweep field projections over $> 100\ \upmu$T. Finally, we present the first application of the frequency-locking technique to an NV ensemble and perform measurements along all four crystallographic axes in rapid sequence to reconstruct the full magnetic field vector. This demonstration of robustness against non-magnetic noise, high dynamic range, and vector capability in a single magnetic sensing device marks a significant advance in transitioning lab-based systems to practical sensors operating in situations with limited control over environmental factors such as vibration, temperature, and ambient magnetic field.

\begin{figure}[]
\includegraphics[width=3.1in]{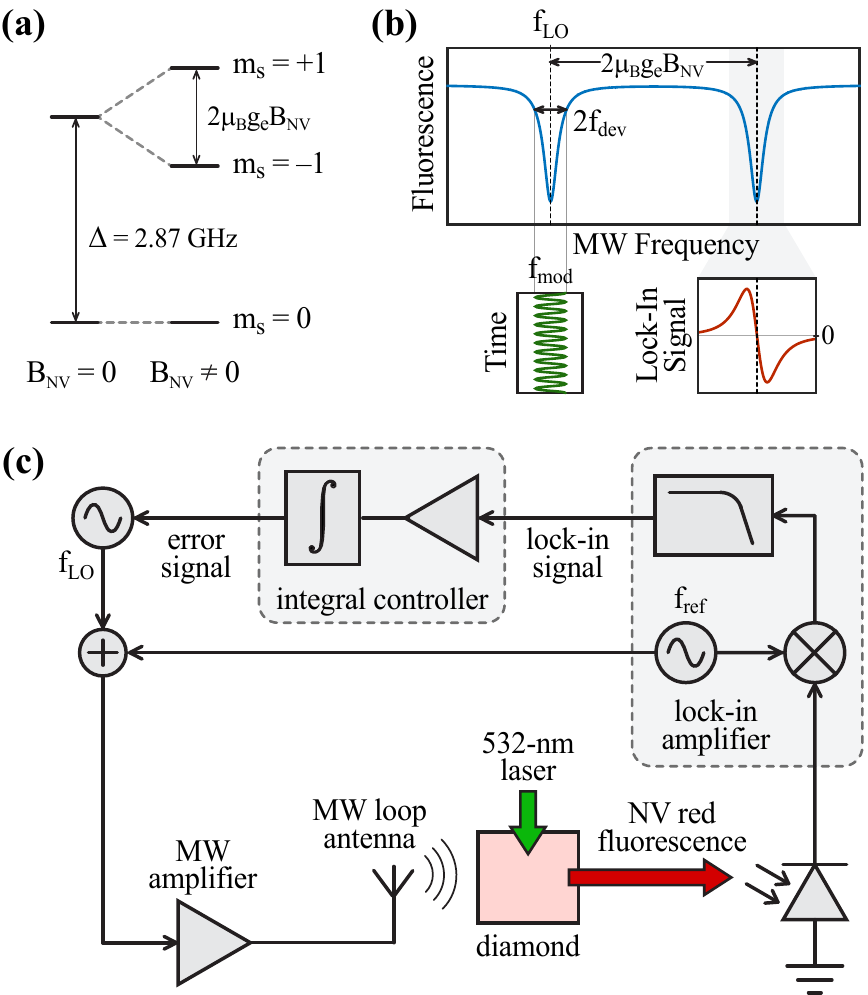}
\caption{ Simplified electronic state energy level diagram (a) and optically detected magnetic resonance spectrum (b) of an NV electronic spin in the presence of an applied magnetic field $B_{\text{NV}}$. The microwave signal $f_{\text{mod}}$ is generated by modulating $f_{\text{LO}}$ at frequency $f_{\text{ref}}$ with depth $f_{\text{dev}}$. The demodulated lock-in signal is shown for $f_{\text{LO}}$ near an NV resonance, illustrating that the lock-in signal is zero when $f_{\text{LO}}$ is directly on resonance. 
(c) Signal path for a single frequency locked loop.
\label{diagram}
}
\end{figure}

\section{Background}
The operating principle of the NV-based magnetometer~\cite{Taylor2008} relies on the the spin-triplet ground state of the NV fine structure, as illustrated in Fig.~\ref{diagram}(a). Under no external magnetic field, spin-spin interactions shift the $m_s=\pm1$ sublevels from the $m_s = 0$ sublevel by the zero field splitting $\Delta \approx 2.87 \text{ GHz}$ at room temperature~\cite{Doherty20131}. In the presence of an external magnetic field, the $m_s = \pm1$ sublevels are shifted by the Zeeman effect. In the limit where the component of the magnetic field transverse to the NV symmetry axis is negligible, the frequencies of $m_s=0 \rightarrow \pm1$ transitions are given by \cite{RondinRPP2014}
% checked with Carson -- transverse fields are not considered in the analysis
\begin{equation} \label{frequencyshifts}
f_{0\pm} \approx \Delta + \beta_\text{T} \delta_\text{T} \pm \gamma B_{\text{NV}},
\end{equation}
where $\beta_\text{T} \approx -74 \text{ kHz/K}$ near room temperature \cite{Acosta2010}, $\delta_\text{T}$ is the temperature offset from 300 K during the measurement, $\gamma = g_e \mu_B/h 
\approx 28~\text{Hz/nT}$ is the NV gyromagnetic ratio \cite{RondinRPP2014}, and $B_{\text{NV}}$ is the projection of the applied magnetic field along the NV symmetry axis.

A common method for sensing magnetic fields with NV centers relies on monitoring the optically detected magnetic resonance (ODMR) spectrum, measured by applying continuous above-band optical excitation and simultaneously sweeping the frequency of an applied microwave (MW) field. The optical excitation polarizes the NV center into the more fluorescent $m_s = 0$ spin state, and when the MW field is on resonance with an $m_s=0 \rightarrow \pm1$ transition, there is a decrease in the fluorescence intensity that manifests as a dip in the ODMR spectrum [Fig. \ref{diagram}(b)]. By simultaneously measuring both $m_s=0 \rightarrow \pm1$ transition frequencies, the effects of temperature and magnetic field can be decoupled, as seen in Eq. \ref{frequencyshifts}. 
This approach can be extended to achieve vector magnetometry by measuring the ODMR spectrum of an NV ensemble to determine the resonance frequencies of all four possible NV axes. In this case, it is necessary to consider additional shifts in the NV resonances due to non-negligible transverse magnetic field components in order to properly reconstruct the full vector magnetic field.

Previous demonstrations of ODMR-based NV magnetometry measured full spectra and performed fits to extract the NV resonance frequencies \cite{SteinertRSI2010, MaertzAPL2010}. While this method is straightforward and able to access the full $> 10$ mT dynamic range of the NV center, it also involves spending a large fraction of the measurement time monitoring non-information-containing, off-resonance signal and is subsequently prohibitively slow for some applications. Fitting the curves also introduces latency that may be incompatible with real-time sensing.
More recent demonstrations used lock-in techniques to continuously monitor a single resonance on the approximately-linear derivative section of the spectral feature, from which small resonance frequency shifts were detected by applying pre-calibrated scale factors \cite{ShinJAP2012, Clevenson2015, barry2016optical}. However, this second approach is limited to the approximately linear regime of the lock-in signal, resulting in a dynamic range of a few $\upmu$T. Furthermore, this method is inherently dependent on phenomenological variables instead of a true frequency shift. In particular, the scale factor is influenced by the NV resonance linewidth and contrast, both of which are affected by optical pump power, microwave power, and detection efficiency \cite{Jensen2013}. These variables are different for each device and will also drift over time, consequently requiring periodic recalibration.

\section{Frequency-Locking Technique}

Here we present a closed-loop system that directly locks the microwave drive field to one or more NV resonances, thus isolating the magnetic field measurement from the phenomenological variables that determine the scale factor of previous lock-in-based approaches. Furthermore, by virtue of locking to the NV resonance frequency, the measurement is always performed in the approximately linear regime of the lock-in signal, enabling maximal sensitivity over the full dynamic range of the NV center. This frequency-locking technique is similar to that employed canonically in atomic $M_z$ type magnetometers~\cite{Alexandrov1992} and, more recently, demonstrated on single NV centers in diamond~\cite{SchoenfeldPRL2011,GrinoldsNatNano2014}. We extend the technique to perform measurements on multiple NV resonances simultaneously, as is necessary to fully decouple temperature from magnetic field, and apply it to an NV ensemble to extract the full magnetic field vector, thus demonstrating a capability not inherent to $M_z$ type atomic magnetometers nor single NV centers.

Figure \ref{diagram}(c) shows a diagram of the signal path of a frequency locked loop for a single NV resonance. Note that while the single-channel case is discussed here for simplicity, simultaneous monitoring of $n>1$ NV resonances is achieved straightforwardly by adding additional frequency channels.
A signal generator outputs a microwave signal frequency, $f_{\textrm{LO}}$, which is tuned to minimize the error on the feedback compensator. The microwave signal is modulated at frequency $f_{\textrm{ref}}$ with depth $f_{\textrm{dev}}$ to produce a time-dependent frequency $f_{\textrm{mod}}(t)$ given by:
\begin{equation} \label{modulation}
f_{\textrm{mod}}(t)=f_{\textrm{LO}}+f_{\textrm{dev}} \cos (2 \pi f_{\textrm{ref}} t).
\end{equation}
The signal is sent to an antenna which drives the NV spin ensemble with the modulated microwave field. The fluorescence signal from the diamond sample is detected with a photodetector and demodulated by a lock-in amplifier. 
We set the phase of the lock-in amplifier such that the in-phase component of the lock-in signal is positive (negative) if $f_{\textrm{LO}}$ is slightly below (above) the NV transition frequency, whereas the lock-in signal is zero when $f_{\textrm{LO}}$ is directly on resonance with the NV transition frequency [see Fig. \ref{diagram}(b)]. A feedback compensator, consisting of a discrete integrator controller, produces an error signal from the lock-in signal. The error signal in turn provides feedback to adjust the microwave source frequency $f_{\textrm{LO}}$ and lock it to the center of the NV resonance.
In an $n$-frequency-channel implementation of this technique, each NV resonance is modulated at a different frequency $f_{\textrm{ref},n}$ to allow for demodulation of all channels simultaneously. In this way, individual lock-in signals and subsequent error signals are extracted for each frequency channel concurrently.

The dynamics of the feedback loop can be analyzed using a linear approximation of the system (see \cite{supplemental} for detailed derivation), such that the plant transfer function is given by
\begin{equation} \label{linearplant1}
G(z)=\frac{2 C V_0}{\sigma^2} f_{\textrm{dev}} \frac{\alpha z}{z + \alpha -1}.
\end{equation}
From Eq.~\ref{linearplant1}, we can see that when this system is applied to open-loop measurements, the gain is proportional to a number of experimental parameters: signal contrast ($C$), fluorescence intensity ($V_0$), and resonance linewidth ($\sigma$). These parameters are susceptible to fluctuations in both laser intensity and microwave power, which degrade the accuracy of the open-loop sensor.
The addition of a feedback loop allows for tracking the frequency of the resonance without this dependence on non-fundamental experimental parameters. Employing a digital integrator controller with gain $K_I$, the resulting closed loop transfer function is given by
\begin{equation} \label{closedlooptransfer}
T(z)=\frac{G(z) C(z)}{1+G(z) C(z)},
\end{equation}
where $C(z)=(K_I z)/(z-1)$ is the transfer function of the digital integrator controller. For large gain values, the closed loop transfer function is approximately 
independent of experimental parameters and is thus robust against laser and microwave intensity noise. Furthermore, invoking the final value theorem with a step input, Eq. \ref{closedlooptransfer} can be shown to have no steady-state error.  The closed loop system accurately tracks the resonant frequency in steady-state, and only the dynamics of the transient response are influenced by non-fundamental experimental parameters.

\begin{figure}[]
\includegraphics[width=3.1in]{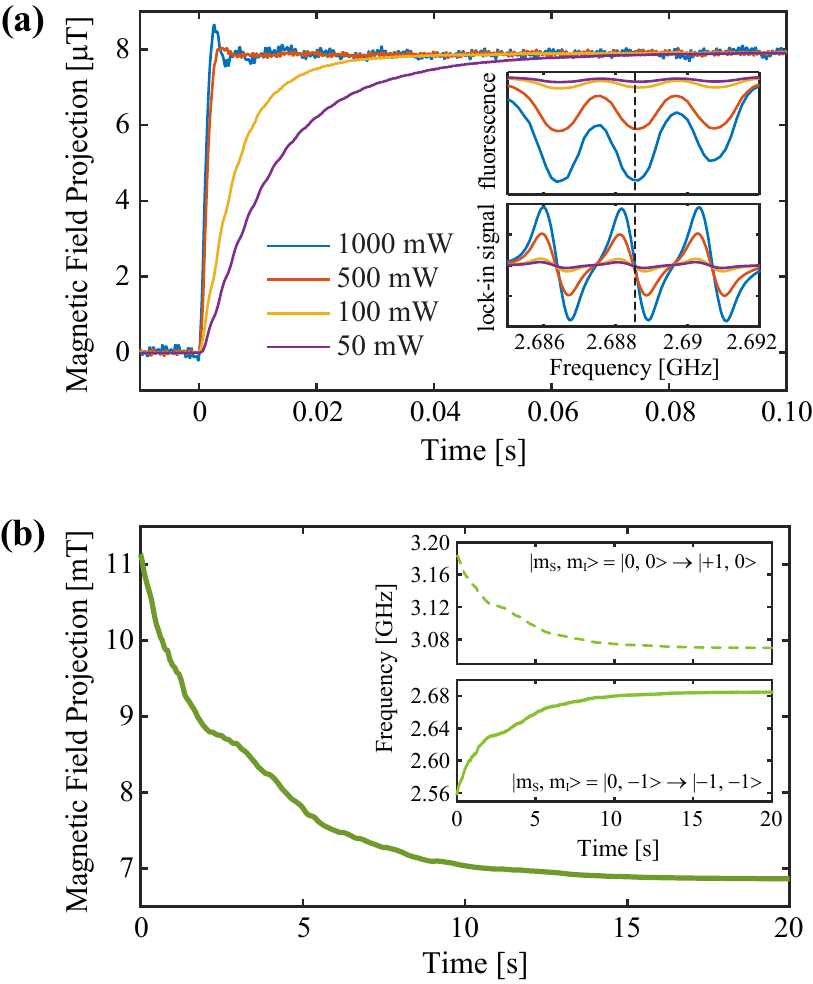}
\caption{
(a) Magnetic field step response of Zeeman splitting showing that the same steady-state magnetic field is measured, independent of large differences in laser power.  Upper inset: ODMR spectra of a single NV resonance.  Lower inset: ODMR spectra at the output of the lock-in amplifier. The dashed vertical lines indicate the locked hyperfine transition.
Note that changes in the laser power affect the contrast as well as the slope of the lock-in signal, degrading the accuracy of an open-loop measurement. In comparison, the steady-state frequency determined by the closed-loop measurement is consistent across the wide range of laser powers.
(b) Dynamic range demonstration of the magnetic field produced by a free-moving permanent magnet varying projection by 4 mT during a continuous acquisition. Simultaneous measurements of the upper and lower resonance frequency of a single NV orientation class are shown in the insets.
} \label{step}
\end{figure}

\section{Experimental Demonstration}

We built a dual-channel, frequency-locking NV magnetometer using a field-programmable gate array (FPGA) and a high-speed digital-to-analog-converter to digitally synthesize two carrier signals with frequencies that can be independently tuned and modulated. The generated signal is amplified and sent to a loop antenna to produce a microwave field at the diamond sample, which contains a large ensemble of NV centers ($\sim 1 \times 10^{12}$). The diamond sample is optically excited with a 532-nm laser 
in a light-trapping diamond waveguide geometry for increased optical excitation efficiency \cite{Clevenson2015}. The resulting fluorescence is collected with a balanced photodetector 
 where a pick-off from the laser is directed into the balancing port to remove common-mode laser noise. The lock-in amplifiers for demodulating the balanced photodetector signal and feedback compensators for locking to the NV resonances are also programmed into the FPGA, which continuously outputs the two locked frequencies to a computer.

Permanent magnets arranged in a Halbach array configuration produce a magnetic bias field $B_0 \approx 7.8$~mT, with $<$0.1\% uniformity across the diamond. The orientation of the bias field with respect to the diamond axes serves to spectrally separate the resonances of the four NV classes. The dual-channel frequency-locking NV magnetometer then simultaneously locks to both the $m_s = 0 \rightarrow \pm 1$ transitions of a single NV orientation in order to decouple the effects of temperature and magnetic field, and thus extract the magnetic field projection along the NV symmetry axis. To avoid depopulation and first order coherent driving effects, which affect the line shape \cite{holes}, different nuclear spin states are addressed for the upper and lower transitions. 

Figure \ref{step}(a) demonstrates the robustness of the frequency-locking NV magnetometer against changes in phenomenological variables, such as optical pump and microwave drive power, which affect the contrast and linewidth of the ODMR spectra. We frequency lock to the $m_s = 0 \rightarrow \pm 1$ transitions of the NV orientation class that is most aligned to the applied magnetic field, using modulation frequencies $f_{\textrm{ref},1} = 1.8240$~kHz and $f_{\textrm{ref},2} = 2.2813$~kHz and modulation depths $f_{\textrm{dev},1,2} = 320$~kHz. Note that the sensitivity degrades at modulation frequencies $>$20 kHz, due to limitations in the NV response bandwidth \cite{ShinJAP2012, barry2016optical}. 
 By varying the optical pump power or microwave amplitude, the contrast of the ODMR resonance (upper inset) varies by more than an order of magnitude, which correspond to varying slopes in the lock-in signal (lower inset). In previous open-loop measurement techniques \cite{ShinJAP2012, Clevenson2015, barry2016optical}, this change in scale factor would result in a corresponding systematic error in magnetic field. 
 Using the frequency-locking technique, while large variations in the optical pump power result in varying transient responses to a step input as discussed previously, the measured steady-state magnetic field remains consistent. The laser power variations shown in Fig. \ref{step}(a)  illustrate the robust nature of the frequency-locking method although practical RF and intensity fluctuations are smaller. 
In open-loop implementations, for example, a $\sim 5\%$ drift in laser intensity introduces an additional $\sim 5\%$ systematic error in the measured magnetic field, thus degrading the accuracy of an open-loop compared to a closed-loop NV magnetometer. 

We employ a free-moving rare-earth permanent magnet to vary the magnetic field experienced by the frequency-locking NV magnetometer [see Fig. \ref{step}(b)], thus demonstrating a dynamic range $\sim 4$ mT in addition to the $\sim 7$ mT bias field. 
A combination of the limited spatial access in the experimental apparatus and the need for a uniform magnetic field across the diamond in order to avoid excessive broadening of the NV resonances constrained the permanent magnet in both size and proximity to the diamond, thus limiting the maximum measured field to $\sim 11$ mT.
In contrast, in an open-loop lock-in based implementation, the dynamic range is limited to the linear region of the ODMR derivative signal, $\sim 2 h \sigma/(10 g_e \mu_B)$, where $\sigma$ is half of the linewidth of the NV resonance. For a typical $\sigma \approx 0.5$ MHz, the dynamic range is $\sim 4\ \upmu$T, three orders of magnitude less than the range demonstrated here. It is important to note that this increased dynamic range is not obtained at the expense of sensitivity, which we measure to be $\sim 1~\textrm{nT}/\sqrt{\text{Hz}}$~\cite{supplemental}. 
While a dynamic range $\sim 4$ mT was demonstrated here, in general the dynamic range of the frequency-locking NV magnetometer is limited by the ambiguity when resonances of different NV orientation classes cross each other, which can be overcome by simultaneously detecting along all four possible diamond tetrahedral directions and performing real-time analysis to take advantage of the redundancy of the overdetermined system, and ultimately by fluorescence suppression near the ground- and excited-stated level anti-crossings around $\sim 50$ mT and $\sim 100$ mT \cite{Epstein2005}. 

\begin{figure}
\includegraphics[width=3.25in]{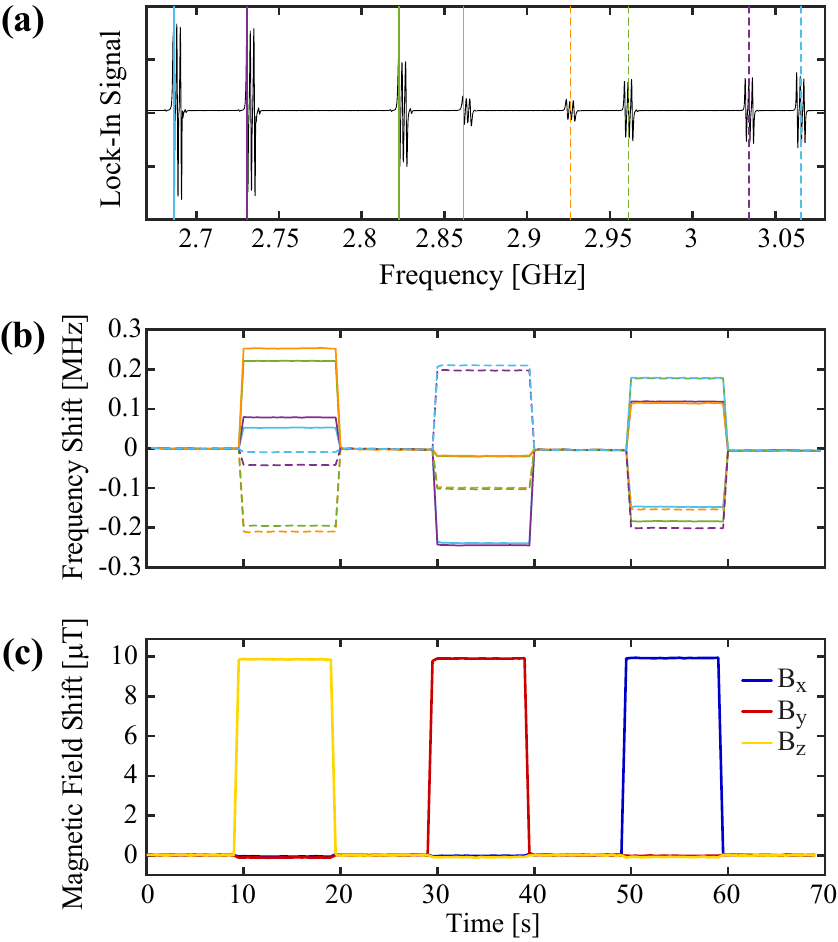}
\caption{
(a) Lock-in ODMR spectrum showing all eight NV resonances. The colored vertical lines indicate the hyperfine transitions chosen for frequency-locking, with each different color corresponding to a single NV orientation class. Solid and dashed lines of the same color were locked simultaneously.
(b) Frequency shifts of all eight NV resonances as a $10\ \upmu$T field was applied sequentially in three orthogonal directions.
(c) Magnetic field components reconstructed from the frequency shift data shown in (b).
} \label{vector}
\end{figure}

Figure \ref{vector} illustrates the capability of the the dual-channel frequency-locking NV magnetometer to measure vector magnetic fields. As before, a Halbach array of permanent magnets applies a uniform magnetic field over the NV ensemble within the diamond sample, causing Zeeman splitting in the $m_s = 0 \rightarrow \pm1$ transitions of the four NV orientation classes and producing eight distinct NV resonances, each with 3 hyperfine transitions [Fig. \ref{vector}(a)]. We simultaneously lock to the $|m_s, m_I \rangle = |0,-1 \rangle \rightarrow |-1,-1 \rangle$ and $|0,0 \rangle \rightarrow |+1,0 \rangle$ hyperfine transitions of a single NV orientation class (corresponding to the thin color-coordinated solid and dashed lines, respectively) and sequentially iterate through each NV orientation, with a 0.1-second dwell time per resonance pair.  A set of three-axis Helmholtz coils apply additional $10\ \upmu$T magnetic fields along three orthogonal directions. Figure \ref{vector}(b) shows the resulting NV resonance frequency shifts detected using this method of locking to four pairs of NV transitions in rapid sequence. Using Eq. \ref{frequencyshifts} and the known, rigid tetrahedral geometry of diamond crystal, we are able to reconstruct the magnetic field vector by performing nonlinear optimization of the overdetermined system, which results from having measurements of the magnetic field projections along four directions rather than three. The reconstructed magnetic field vector is presented in Figure \ref{vector}(c) and is in excellent agreement with the expected magnetic field produced by the calibrated three-axis coils. Note that while we demonstrated full vector reconstruction by rapidly interleaving detection along each of the four possible NV orientation classes, simultaneous magnetic vector measurements can also be performed by extending the frequency-locking NV magnetometer's capability from two frequency channels to eight. 

\section{Summary}

The nitrogen-vacancy center in diamond possesses a wide range of properties that make it exceptionally suitable as a magnetometer. In particular, the NV system benefits from an atom-like electronic structure with the long-term stability afforded by magnetic field measurements being inherently tied only to fundamental constants~\cite{Taylor2008,Maze2008,Balasubramanian2008}; long room-temperature coherence times allowing for sensitive measurements under ambient conditions~\cite{Maze2008,BalasubramanianNatMat2009}; a well-characterized temperature dependence enabling operation over an extreme range of temperatures~\cite{Acosta2010}; inherently high dynamic range; vector capability tied intrinsically to the stable diamond lattice~\cite{SteinertRSI2010, MaertzAPL2010, PhamNJP2011}; and a flexible solid-state geometry that supports measurement modalities ranging from super-resolution magnetic imaging~\cite{BalasubramanianNature2008, Maletinsky2012, RondinAPL2012, JaskulaOpticsExpress2017} to bulk magnetometry with a high-density, compact sensor~\cite{Acosta2010a, Clevenson2015, Barry2016}. The combination of these capabilities in a single device significantly expands the functionality of magnetic sensors, potentially transforming the magnetic sensing application space, especially in field situations where there is limited control over environmental factors~\cite{Cui201570, Shockley:2014uq, CancianiIEEE2017, foote1996relationship, kvamme2006magnetometry, espy2010ultra, sudac2009detection, mcfee1994total, NaraIEEE2006, Holzinger2014}. 

In this work, we have addressed several key technical issues in order to realize many of these inherent capabilities in a single implemented device. We have demonstrated a multi-channel, simultaneous frequency-locking technique for scale-factor-free magnetic field measurements that are robust against temperature fluctuations, achieve high dynamic range, and are applicable to an NV ensemble for full reconstruction of the magnetic field vector. Such a demonstration marks a vital advance in transitioning the NV magnetometer from a laboratory system to a functional device for detecting fields in a practical environment.

\section*{Acknowledgments}
The authors would like to thank J. F. Barry, C. McNally, and M. Walsh for helpful discussion.  The Lincoln Laboratory portion of this work is sponsored by the Assistant Secretary of Defense for Research \& Engineering under Air Force Contract \#FA8721-05-C-0002 and the Office of Naval Research Section 321MS. Opinions, interpretations, conclusions and recommendations are those of the authors and are not necessarily endorsed by the United States Government. H.\,C. was supported by the NASA Office of the Chief Technologist's Space Technology Research Fellowship and the MIT LL Integrated Quantum Initiative. D.E. acknowledges partial support from the AFOSR PECASE (supervised by Dr. Gernot Pomrenke). \\

\bibliography{FrequencyLocking}

\end{document}